# Experimental study of fusion neutron and proton yields produced by petawatt-laser-irradiated $D_2$-$^3$He or $CD_4$-$^3$He clustering gases

W. Bang,[1, a)] M. Barbui,[2, b)] A. Bonasera,[2, 3] H. J. Quevedo,[1] G. Dyer,[1] A. C. Bernstein,[1] K. Hagel,[2] K. Schmidt,[2] E. Gaul,[1] M. E. Donovan,[1] F. Consoli,[4] R. De Angelis,[4] P. Andreoli,[4] M. Barbarino,[2] S. Kimura,[5] M. Mazzocco,[6] J. B. Natowitz,[2] and T. Ditmire[1]

[1]*Center for High Energy Density Science, C1510, University of Texas at Austin, Austin, TX, 78712, USA*
[2]*Cyclotron Institute, Texas A&M University, College Station, TX, 77843, USA*
[3]*LNS-INFN, V.S.Sofia 64, 95123 Catania, Italy*
[4]*Associazione Euratom - ENEA sulla Fusione, via E. Fermi 45, CP 65-00044 Frascati (Rome), Italy*
[5]*Department of Physics, University of Milano, via Celoria 16, 20133 Milano, Italy*
[6]*Department of Physics and Astronomy and INFN-Sezione di Padova, via F. Marzolo 8, I-35131 Padova, Italy*

**Abstract**
We report on experiments in which the Texas Petawatt laser irradiated a mixture of deuterium or deuterated methane clusters and helium-3 gas, generating three types of nuclear fusion reactions: $D(d, ^3He)n$, $D(d, t)p$ and $^3He(d, p)^4He$. We measured the yields of fusion neutrons and protons from these reactions and found them to agree with yields based on a simple cylindrical plasma model using known cross sections and measured plasma parameters. Within our measurement errors, the fusion products were isotropically distributed. Plasma temperatures, important for the cross sections, were determined by two independent methods: (1) deuterium ion time-of-flight, and (2) utilizing the ratio of neutron yield to proton yield from $D(d, ^3He)n$ and $^3He(d, p)^4He$ reactions, respectively. This experiment produced the highest ion temperature ever achieved with laser-irradiated deuterium clusters.

Authors to whom correspondence should be addressed. Electronic mail:
a) dws223@physics.utexas.edu.
b) barbui@comp.tamu.edu

## I. Introduction

Dynamics of nuclear fusion reactions generated by the Coulomb explosion of laser-heated molecular clusters have been actively studied both experimentally and theoretically for over a decade [1-13]. In those studies, an intense laser pulse irradiates either cryogenically cooled deuterium ($D_2$) or near room temperature deuterated methane ($CD_4$) cluster targets. The interaction between the laser pulse and the clusters leads to an explosion of the clusters generating multi-keV ions, a process successfully explained by the so-called Coulomb explosion model [1-5,14]. This model applies when the clusters are almost completely ionized by a strong laser field in a very short time compared with the ion motion [1,15-18]. In this case, the highly charged clusters of deuterium ions promptly explode by Coulomb repulsion creating a hot plasma. The resultant deuterium ions are energetic enough to fuse within the cluster jet, and the fusion plasma produces a burst of 2.45 MeV neutrons and 3 MeV protons among other fusion products.

One driving goal of these cluster fusion studies is to develop high flux neutron pulses for time-resolved neutron damage studies, which require neutron yields greater than $10^9$–$10^{10}$ neutrons per shot [13,19]. Since the DD fusion cross-section increases very quickly with the ion temperature in the 1–30 keV range [20], accessible in laser-cluster fusion experiments, it is very important to determine the plasma temperature accurately. To measure the ion temperature, researchers have often used time-of-flight (TOF) diagnostics [11,21]. Although it is not obvious that the temperature measured from the ion TOF data is the same as the plasma temperature at the time of the fusion reactions, we validated in a recent experiment that this is indeed the case [22]. In that work, as in this one, an intense ultrashort laser pulse irradiated either $D_2$ clusters mixed with helium-3 gas or $CD_4$ clusters mixed with helium-3 gas, simultaneously producing three types of nuclear fusion reactions in the interaction volume: $D(d, {}^3He)n$, $D(d, t)p$, and ${}^3He(d, p){}^4He$. The ratio of the 2.45 MeV neutron yield from the D(d,

$^3$He)n reaction to the 14.7 MeV proton yield from the $^3$He(d, p)$^4$He fusion reaction directly determines the temperature of the fusion plasma.

In this paper, we study the yields of fusion neutrons and protons from those reactions. First, we present data obtained with deuterium clusters only, and show that the 2.45 MeV neutron and 3 MeV proton yields agree with each other, as expected. Adding $^3$He gas in the cluster targets, we measured the yields of 2.45 MeV neutrons and 14.7 MeV protons and examined the angular distribution of these particles. In order to investigate whether our current fusion model, presented in Section III of this paper, can accurately predict how many neutrons and protons are produced from a given fusion plasma, we compared the measured fusion yields with the fusion yields as predicted by our model.

## II. Experimental setup

The Texas Petawatt laser (TPW) [23] delivered 90–180 J per pulse with 150–270 fs duration to irradiate the clusters. An f/40 focusing mirror (10 m focal length) created a large interaction volume in this experiment with laser intensities sufficiently high to drive cluster fusion reactions. With this focusing geometry, a neutron yield as high as $1.6\times10^7$ n/shot was observed previously on the TPW using D$_2$ clusters [10].

We measured the on-shot energy and pulse duration of each shot using the calibrated TPW output sensor package (OSP). We also measured the laser energy that was not absorbed or scattered by the cluster target using an energy meter behind two black glass plates of known transmission. Two cameras imaged the side and bottom of the plasma on each shot while a third camera, located in the OSP, acquired an image of the beam profile at the equivalent plane of the cluster target, allowing us to estimate the radius, *r*, of the cylindrical fusion plasma. The measured diameter of the laser pulse was consistent with the diameter of the plasma observed using the side and bottom images of the plasma.

A mixture of $D_2$ or $CD_4$ clusters and $^3$He gas served as the target. A residual gas analyzer measured the partial pressures of $D_2$, $CD_4$, and $^3$He in the mixture, from which we calculated the ratio of the atomic number densities of deuterium and $^3$He for each shot. The gas mixtures were introduced at a pressure of 52.5 bars into a conical supersonic nozzle with a throat diameter of 790 μm, an exit radius of $R$ =2.5 mm, and a half angle of 5 degrees to generate large clusters (>10 nm) necessary for energetic cluster explosions. In order to produce large clusters, we cooled the $D_2 + {}^3$He and $CD_4 + {}^3$He mixtures to 86 K and 200-260 K, respectively. The $CD_4$ gas was flowed at a temperature above 200 K to prevent it from condensing. The cooled $CD_4$ clusters are expected to be 15–40% larger than those at room temperature, whereas $D_2$ gas does not form clusters in significant numbers at room temperature [24,25]. A series of Rayleigh scattering measurements confirmed that the average size of $D_2$ clusters remained largely unchanged (less than 7% decrease in diameter) when $^4$He gas was added, and we assume the same with the addition of $^3$He into $D_2$ under the same conditions.

A Faraday cup, located 1.07 m from the plasma with an opening diameter of 16 mm, provided the TOF measurements of the energetic deuterium and carbon ions arriving from the plasma. A ground mesh placed in front of the cup maintained a field-free region near the Faraday cup, while a negative 400 V bias on the collector repelled the slow electrons that could affect the TOF measurements arriving at the same time as the ions. We calculated the total number and energy spectrum of deuterium ions in the laser plasma, with the assumption of isotropic emission [10,21]. This assumption is reasonable because the clusters undergo Coulomb explosion in this experiment rather than the ambipolar expansion according to the criteria in Ref. [4].

Five calibrated plastic scintillation detectors [26] measured the yield of 2.45 MeV neutrons generated from DD fusion reactions: three EJ-232Q plastic scintillation detectors measured the neutron yield at 1.9 m from the plasma, while the other two EJ-200 plastic scintillation detectors measured the yield at 5 m to increase the dynamic range. To investigate the angular distribution of the neutron yield, four calibrated NE213 liquid scintillation detectors measured the neutron yield at 36º, 90º, and 151º with respect to the laser propagation direction.

Three plastic scintillation detectors measured the yield of 14.7 MeV protons from the $^3$He(d,p)$^4$He fusion reactions. These proton detectors were located in vacuum 1.06–1.20 m from the plasma at 45º, 90º, and 135º. Calibration of the detectors was performed prior to the experiment at the Cyclotron Institute, Texas A&M University, using 14.7 MeV proton beams delivered by the K150 Cyclotron. Each proton detector consists of a 41.8 mm diameter, 0.254 mm thick BC-400 scintillator and two photomultiplier tubes on the left and right side of the scintillator. A 1.10 mm thick aluminum degrader was inserted in front of each detector to block all the other charged particles including the 3 MeV protons originating from DD fusion reactions, and to slow the 14.7 MeV protons to 4.0 MeV so that they could transfer all of their remaining kinetic energy to the scintillator disk. When proton detectors operated with 25 μm thick aluminum degraders, they also detected the 3 MeV protons from DD fusion reactions, but the degraders still completely blocked 1.01 MeV tritium and 0.82 MeV $^3$He ions as well as the deuterium ions coming from the plasma. A Monte Carlo simulation code, SRIM [27], calculated the energy loss of deuterium ions and fusion products in various materials to aid in the design of the degraders.

## III. Cluster fusion model with a mixture target

Figure 1 illustrates some of the possible fusion reactions within the gas jet when both $D_2$ clusters and $^3$He atoms are present. A focused intense laser beam, which has a boundary represented by a black circle, illuminates the target and is strongly absorbed. The deuterium clusters, portrayed much larger than actual scale in the figure, undergo Coulomb explosion, and the kinetic energies of the resulting deuterium ions reach several keV. These ions can collide with each other and generate DD fusion reactions (which we call beam-beam fusion). DD fusion can also occur when energetic deuterium ions collide with cold deuterium atoms in the background gas jet outside the focal spot (beam-target fusion). Since there are also ³He atoms or ions present in the gas jet, ³He(d, p)⁴He fusion reactions can happen. Unlike deuterium ions undergoing Coulomb explosion, the ³He atoms do not absorb the laser pulse energy efficiently because they do not form clusters at 86 K [28]. Therefore, the ³He ions remain cold after the intense laser pulse is gone, and we consider ³He ions stationary for purposes of our fusion modeling. In our model, the "plasma temperature" refers to the temperature of deuterium ions only.

With both energetic deuterium ions from explosion of the clusters and cold ³He ions in the background gas jet, some of the possible fusion reactions inside the plasma are:

D + D → T (1.01 MeV) + p (3.02 MeV) (50%),  (1a)

D + D → ³He (0.82 MeV) + n (2.45 MeV) (50%),  (1b)

D + ³He → ⁴He (3.6 MeV) + p (14.69 MeV) (100%),  (1c)

all of which we observe with the neutron and proton detectors. Among these reactions, we are particularly interested in (1b) and (1c), from which 2.45 MeV neutrons and 14.7 MeV protons are produced, respectively. Since the two reactions have different cross-sectional dependence on the plasma temperature, the ratio of their yields can uniquely determine the plasma temperature at the critical moment when the fusion reactions occur.

In this experiment, the diameter of the incident laser pulse on the gas jet target is over 1 mm, and the fusion plasmas are approximately cylindrical in shape. This is confirmed by side and bottom images of the plasmas, and allows us to calculate expected fusion yields using the cylindrical plasma model that we developed previously [22,29,30]. The 14.7 MeV proton yield, 2.45 MeV neutron yield, and their density-weighted ratio are given respectively by [2,7,22]:

$$Y_p = N_{ion} n_{3_{He}} <\sigma_{D^3He}>_{\frac{3}{5}kT} R,$$  (2a)

$$Y_n = N_{ion} n_D [\frac{1}{2} <\sigma_{D(d,n)^3He} v>_{kT} \frac{r}{<v>_{kT}} + <\sigma_{D(d,n)^3He}>_{\frac{1}{2}kT} (R-r)],$$  (2b)

$$\frac{Y_p}{Y_n}\frac{n_D}{n_{^3He}} = \frac{\langle\sigma_{D^3He}\rangle_{\frac{3}{5}kT} R}{[\frac{1}{2}\langle\sigma_{D(d,n)^3He}v\rangle_{kT}\frac{r}{\langle v\rangle_{kT}}+\langle\sigma_{D(d,n)^3He}\rangle_{\frac{1}{2}kT}(R-r)]}, \quad (2c)$$

where $N_{ion}$ is the total number of energetic deuterium ions in the plasma, $n_{^3He}$ and $n_D$ are the atomic number densities of $^3$He and deuterium, respectively, $kT$ is the deuterium ion temperature, $\langle\sigma_{D^3He}\rangle_{3kT/5}$ is the average cross-section for D($^3$He, p)$^4$He reactions at the $3kT/5$ center-of-mass temperature (since $^3$He is at rest), $R = 2.5$ mm is the radius of the exit nozzle, $\langle\sigma_{D(d,n)^3He}v\rangle_{kT}$ is the fusion reactivity at $kT$, $r$ is the radius of the cylindrical plasma, $\langle v\rangle_{kT}$ is the mean speed of the hot deuterium ions, and $\langle\sigma_{D(d,n)^3He}\rangle_{kT/2}$ is the average fusion cross-section between hot deuterium ions and cold deuterium atoms. A uniform atomic density was assumed throughout the gas jet for both $^3$He and deuterium. In Eq. (2b), the plasma disassembly time is estimated as $r/\langle v\rangle_{kT}$, and the beam-target contribution, the second term on the right of Eq. (2b), is considered only in the region outside the fusion plasma, over a distance $(R-r)$.

In contrast to the plasma temperature measured in Ref. [22] using Eq. (2c), which does not have explicit dependence on the individual gas jet density or $N_{ion}$, the fusion yields in Eqs. (2a) and (2b) explicitly depend on both the gas jet density of each species and $N_{ion}$. Therefore, comparison of the measured fusion yields and the expected fusion yields from Eqs. (2a) and (2b) provides a stronger test for our cylindrical plasma model.

## IV. Results and analysis

DD fusion reactions produce nearly equal number of protons and neutrons, as indicated in Eqs. 1(a) and 1(b), respectively. Figure 2 is a plot of the 3 MeV proton yield against the 2.45 MeV neutron yield. For this measurement, we used deuterium clusters as the target without adding $^3$He gas. The plotted proton yield is the average value of the two detectors taken for each shot. Similarly, the plotted neutron yield is the average of five plastic scintillation detectors and four liquid scintillation detectors. The error bars indicate one standard deviation of the mean along their respective axis directions. Figure 2 depicts the observed linear relationship between the proton and neutron yields, which demonstrates that the proton detectors worked as designed. Within our measurement errors, the proton yield matched the neutron yield, and we confirmed cross-calibration of the proton and neutron detectors through this measurement.

We also investigated the angular distribution of neutrons and protons using $D_2$ clusters, $D_2$ clusters mixed with $^3$He gas, and $CD_4$ clusters mixed with $^3$He gas. Figures 3(a) and 3(b) show the neutron and proton yields, respectively, at three different angles, normalized to the average of all the detectors over tens of shots under similar experimental conditions. In Fig. 3(a), the hollow black squares, blue circles, and red triangles indicate neutrons from targets of $D_2$ clusters, $D_2 + {}^3$He, and $CD_4 + {}^3$He, respectively. Seven neutron detectors measured the fusion yields at 90 degrees (two liquid and five plastic scintillation detectors), and this resulted in a smaller measurement spread at this angle. The detectors did not show a significant deviation of the angular distribution of neutrons from isotropic within our measurement errors. Based on these results and our knowledge of the physical processes, we assumed isotropic emission of 2.45 MeV neutrons for the neutron yield measurement.

In Fig. 3(b), the solid black squares indicate the yields of 3 MeV protons from reactions in Eq. 1(a) using $D_2$ clusters, whereas the solid blue circles and red triangles indicate the yields of 14.7 MeV protons from reactions in Eq. 1(c) using $D_2 + {}^3$He and $CD_4 + {}^3$He mixtures, respectively. The measured 3 MeV mean proton yield at 45 degrees was nearly identical to the yield at 135 degrees, but the 14.7 MeV mean proton yield measurements showed a deviation up to about 20% from the average yield across our three angles. We attribute this to the statistical uncertainty resulting from having only two proton detectors, and from the low number of protons arriving at each detector. On some shots, there were fewer than five protons per detector. Within experimental uncertainty, the angular distributions of 14.7 MeV protons and 3 MeV neutrons appear to be isotropic as expected. To reduce errors in model-to-experiment yield comparisons involving 14.7 MeV protons, we subsequently used the average yields from the proton detectors at all angles.

Figure 4 is a representative Maxwellian ion speed distribution fit to the ion TOF spectrum, which we used to determine temperature, $kT_{TOF}$. Ions with different kinetic energies arrive at the Faraday cup at different times, and this produces the observed TOF spectrum. The Faraday cup also responds to the strong x-rays originating from the hot electrons generated within the plasma, and displays a strong narrow peak near the time of laser arrival. As the initial x-ray peak decays away, energetic deuterium ions arrive at the Faraday cup and result in the ion TOF signal as shown in Fig. 4. A dashed red line in the figure represents a two-source fit with an exponentially rising and decaying curve accounting for the initial x-rays and a Maxwellian distribution of the energetic ions. In this shot, a laser pulse with an intensity of $\sim 3\times 10^{16}$ W/cm$^2$ irradiated D$_2$ clusters mixed with $^3$He gas, and produced a fusion plasma with $kT_{TOF}$ =18 keV, which is the highest ion temperature ever achieved with laser-irradiated deuterium clusters. Consequently, the fusion plasma produced $1.9\times 10^7$ neutrons on this shot, which is the highest yield achieved in a laser-cluster fusion experiment.

It is important to note that our fusion plasmas are not in thermal equilibrium. The energetic deuterium ions in the plasma have such high kinetic energies that they have longer mean free paths (>10 mm) than the size of the gas jet, and thermalization from ion-ion collisions is not expected under our experimental conditions. Previous studies have shown that near-Maxwellian ion energy distribution is observed not because of thermalization of ions, but because of cluster size distribution [5,21]. A near-Maxwellian energy distribution results because the energy distribution of the plasma is a convolution of the ion energy spectrum owing to the Coulomb explosion of a single cluster with the log-normal size distribution of clusters in our gas jet. Indeed, the ions in this experiment show a near-Maxwellian energy distribution, and this measured ion spectrum determines $kT_{TOF}$.

Using the fusion yield model described in section III, one can calculate the expected proton and neutron yields using Eqs. 2(a) and 2(b), respectively. In this calculation, the measurement of the "correct plasma temperature" plays a crucial role because the fusion cross-sections in Eqs. 2(a) and 2(b) have strong and nonlinear dependences on the plasma temperature. Slightly different temperatures can give rise to very different expected fusion yields.

Figures 5(a) and 5(b) compare the experimentally measured 14.7 MeV proton and 2.45 MeV neutron yields to their expected yields using Eqs. 2(a) and 2(b), respectively. The expected 14.7 MeV proton yield was calculated using Eq. 2(a) and the measured number of energetic deuterium ions, number density of $^3$He, radius of the gas jet, and the fusion cross-section calculated using $kT_{TOF}$. Similarly, we calculated the expected 2.45 MeV neutron yield by inserting the measured values of the plasma parameters into Eq. 2(b). Both figures show good agreements between the measured fusion yields and the calculated yields. Figure 5(b), however, showed less deviation from perfect agreement than Fig. 5(a) owing to the smaller statistical errors in the neutron yield measurements than in the proton yield measurements.

The ion TOF method gives a simple and easy estimate of the plasma temperature. This method, however, has a fundamental limitation: it is not a direct measurement of the ion temperature while fusion occurs inside the gas jet. It is a measurement of the ion energies after the fusion reactions ended and the plasma expanded completely. Any force that lasts longer than the fusion burn time (~1 ns) while the plasma still expands can affect the ion energy spectrum recorded afterwards by the TOF method. Therefore, we need a different method to determine directly the ion temperature of cluster fusion plasmas during the time-period over which actual fusion takes place.

Recently, we measured this temperature, $kT_{Fusion}$ [22], and Figs. 6(a) and 6(b) compare the measured proton and neutron yields with the expected proton and neutron yields, respectively, using $kT_{Fusion}$. The data in Fig. 6(b) were reported in Ref. [22], and are shown here with error bars for a comparison with Fig. 5(b). Both Figs. 6(a) and 6(b) show good agreement between the measured yields and the expected yields although they exhibit slightly more spread compared with Fig. 5(b). The uncertainty in the proton yield measurement, which resulted in the spread in Fig. 5(a), now affects both Figs. 6(a) and 6(b) because the errors in the proton yield measurements propagate as the errors in the density-weighted ratio of fusion yields that determines $kT_{Fusion}$.

The close agreement of measured and predicted yields in Figs. 5(a), 5(b), 6(a), and 6(b) indicates that both $kT_{TOF}$ and $kT_{Fusion}$ are directly correlated with the plasma temperature responsible for the fusion reactions. Although the energy spectrum of ions determine $kT_{TOF}$ several hundred nanoseconds later than the fusion reactions, we find that $kT_{TOF}$ correctly represent the plasma temperature during actual fusion by observing a good agreement between the experimentally measured fusion yields and the expected fusion yields calculated from $kT_{TOF}$. Similarly, the expected fusion yields from $kT_{Fusion}$ agree with the measured fusion yields. If either temperature differed from the temperature during actual fusion reactions, the figures generated using that temperature would have shown large discrepancy instead. This conclusion is consistent with the observation in Ref. [22], where $kT_{Fusion}$ closely matched $kT_{TOF}$.

## V. Conclusions

We experimentally investigated the emission of fusion neutrons and protons from petawatt-laser-irradiated deuterium clusters mixed with $^3$He gas and deuterated methane clusters mixed with $^3$He gas. Fusion yield measurements from three directions strongly supported our working assumption that the emission of 2.45 MeV neutrons or 3 MeV protons from the plasma is nearly isotropic. The emission of 14.7 MeV protons was not definitively different from the isotropic emission. The laser-cluster interaction produced deuterium plasmas with ion temperatures as high as 18 keV, and generated up to $1.9 \times 10^7$ DD fusion neutrons in a single shot. To get the correct plasma temperature, which is critical for getting correct fusion cross sections, we measured ion temperatures by two complementary techniques, TOF and yield ratios, which were in good agreement. We found agreement between experimental and theoretical proton and neutron yields using a simple cylindrical plasma model.

**Acknowledgments**

W. B. would like to acknowledge generous support by the Glenn Focht Memorial Fellowship. This work was supported by NNSA Cooperative Agreement DE-FC52-08NA28512 and the DOE Office of Basic Energy Sciences. The Texas A&M University participation was supported by the US DOE and Robert A. Welch Foundation Grant A0330.

**Figures**

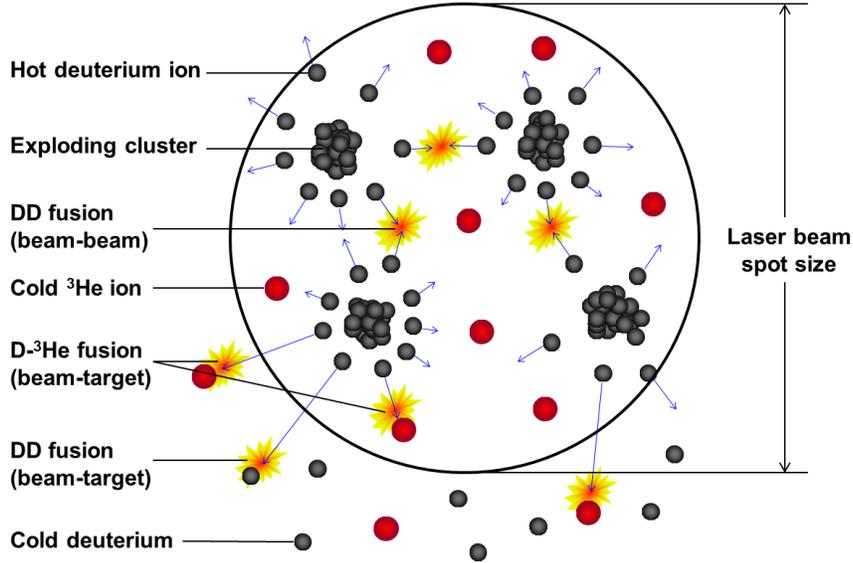

FIG. 1. (Color online) Possible fusion reactions between the constituent particles are shown in this illustration. The bigger red spheres indicate cold $^3$He ions or atoms, and the smaller black spheres represent energetic deuterium ions or cold deuterium atoms.

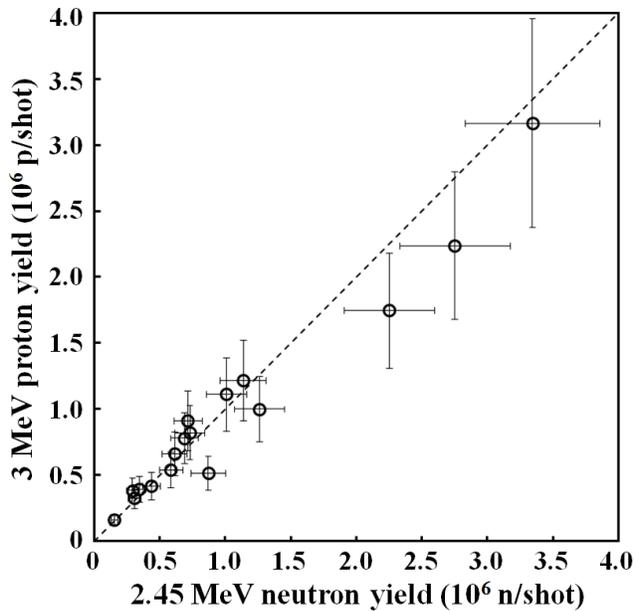

FIG. 2. 3 MeV proton yield vs. 2.45 MeV neutron yield from DD fusion reactions. The straight dashed line indicates when both yields are equal.

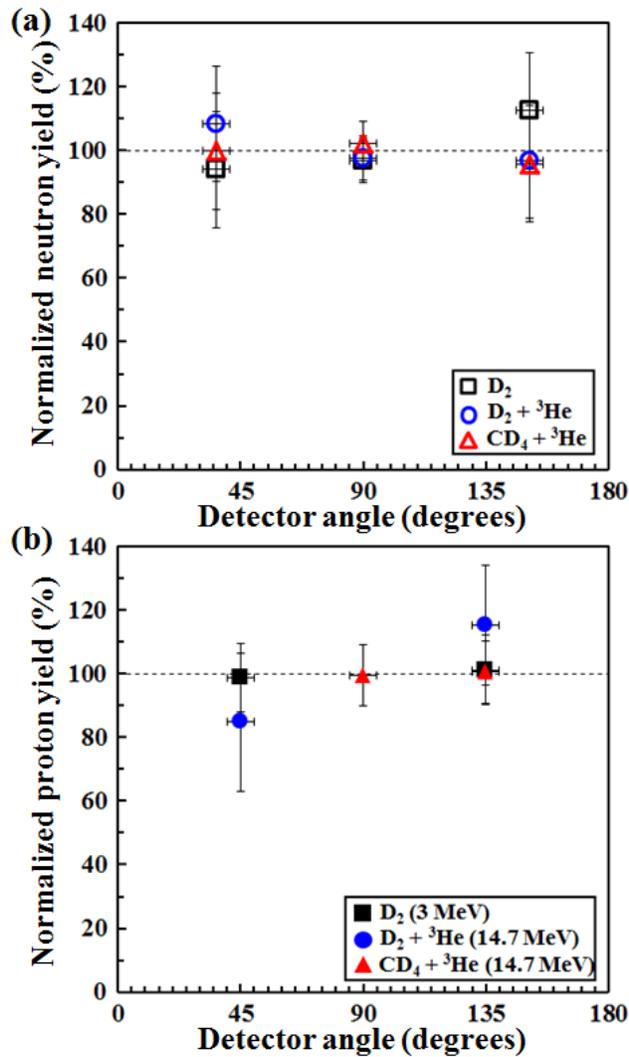

FIG. 3. (Color online) Angular distribution of (a) 2.45 MeV neutrons, (b) 3 MeV protons, and 14.7 MeV protons.

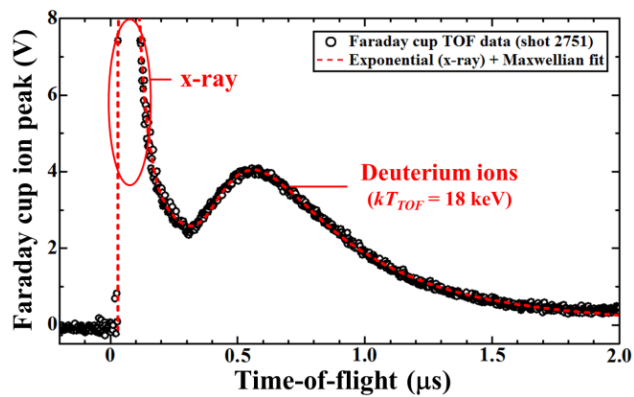

FIG. 4. (Color online) An example of the ion time-of-flight data along with an 18 keV Maxwellian fit and an exponentially rising and decaying curve accounting for the initial x-ray peak near the time of laser arrival (dashed red line).

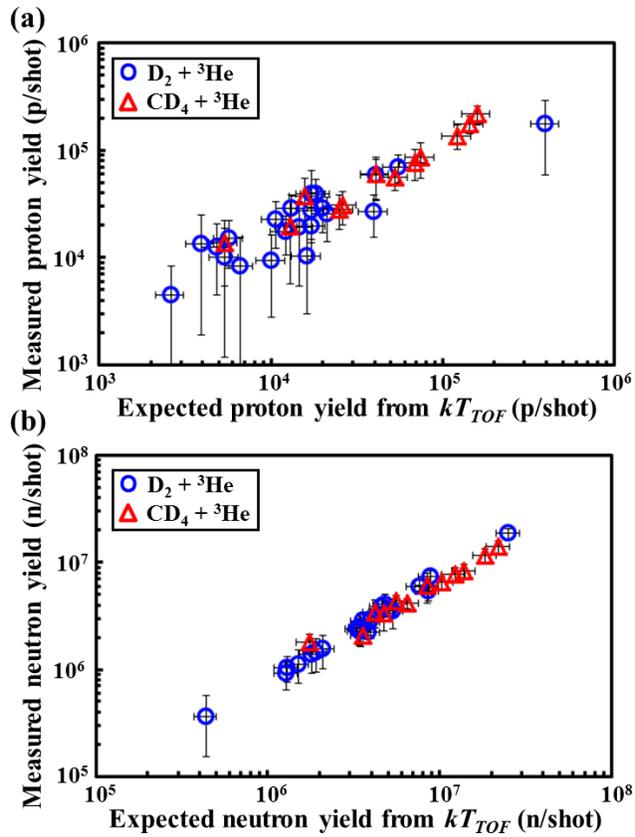

FIG. 5. (Color online) (a) Experimentally measured 14.7 MeV proton yield versus expected proton yield from the model using $kT_{TOF}$. (b) Experimentally measured 2.45 MeV neutron yield versus expected neutron yield from the model using $kT_{TOF}$.

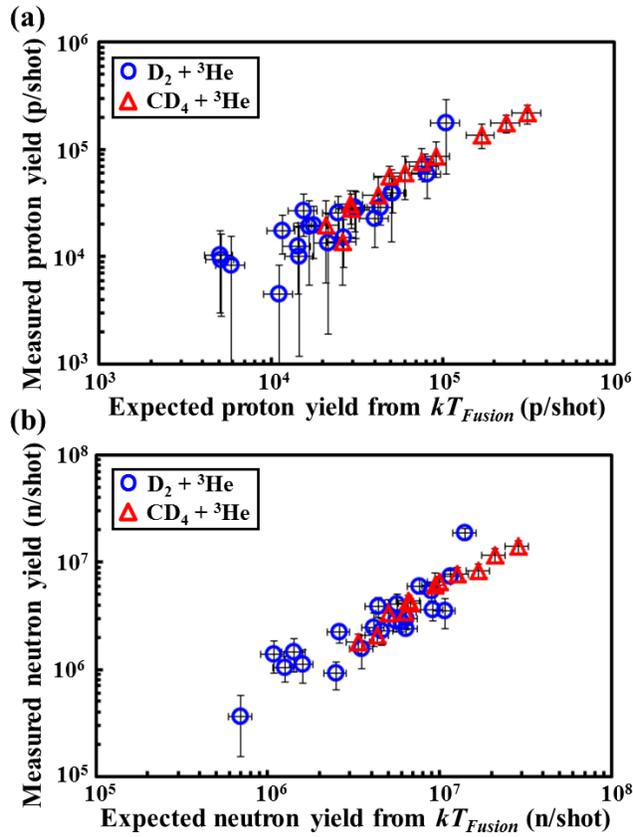

FIG. 6. (Color online) (a) Experimentally measured 14.7 MeV proton yield versus expected proton yield from the model using $kT_{Fusion}$. (b) Experimentally measured 2.45 MeV neutron yield versus expected neutron yield from the model using $kT_{Fusion}$ in Ref. [22].

17